\documentclass[]{emulateapj}
\shorttitle{Runaway in Arches}
\shortauthors{Chatterjee et~al.}

\begin{document}
\title{On the Dynamical Evolution of the Arches Cluster}
\author{Sourav Chatterjee\altaffilmark{1}, Sanghamitra Goswami\altaffilmark{1}, 
Stefan Umbreit\altaffilmark{1}, Evert Glebbeek\altaffilmark{2}, Frederic A.\ Rasio\altaffilmark{1}, and Jarrod Hurley\altaffilmark{3}}
\affil{$^{1}$ Department of Physics and Astronomy, Northwestern University, 
Evanston, IL 60208, USA}
\affil{$^{2}$ Department of Physics and Astronomy, McMaster University, 1280 Main Street 
West, Hamilton, Ontario L8S 4M1, Canada}
\affil{$^{3}$ Center for Astrophysics and Supercomputing, Swinburne University of 
Technology, VIC 3122, Australia}
\begin{abstract}
\label{abstract}
We study the dynamical evolution of the young star cluster Arches and its 
dependence on the assumed initial stellar mass function (IMF). We perform many 
direct $N$-body simulations with various initial conditions and two different 
choices of IMFs. One is a standard Kroupa IMF without any mass segregation.  
The other is a radially dependent IMF, as presently observed in the Arches. 
We find that it is unlikely for the Arches to have attained the 
observed degree of mass segregation at its current age starting from a standard 
non-segregated Kroupa IMF.  We also study the possibility of a 
collisional runaway developing in the first $\sim 2-3\,\rm{Myr}$ of dynamical 
evolution.  We find that the evolution of 
this cluster is dramatically different depending on the choice of IMF: if a 
primordially mass segregated IMF is chosen, a collisional runaway should always 
occur between $2-3\,\rm{Myr}$ for a broad range of initial concentrations. In 
contrast, for a standard Kroupa IMF no collisional runaway is predicted. We 
argue that if Arches was created with a mass segregated IMF similar to what is 
observed today then at the current cluster age a very unusual, high-mass star should be 
created.  However, whether a collisional runaway leads to the formation of an 
intermediate-mass black hole (IMBH) depends strongly on the mass loss 
rate via winds from massive stars.  Growth of stellar mass through collisions 
can be quenched by strong wind mass loss.  In that 
case, the inter-cluster as well as intra-cluster medium are expected to 
have a significant Helium enrichment which 
may be observed via Helium recombination lines.  The excess amount of gas 
lost in winds may also be observed via X-ray observations as diffused X-ray sources.   
\end{abstract}
\keywords{methods: N-body simulations, methods: numerical, globular clusters: individual (Arches), stellar dynamics}
\section{Introduction} 
\label{intro} 
One of the biggest uncertainties in star
cluster evolution studies lies in determining the true initial stellar mass function
(IMF).  Traditionally all studies focusing on the dynamical evolution of star
clusters assume 
an initially fully formed cluster with a given number of single and binary
stars, all at zero age main sequence (ZAMS) at $t=0$ and no gas.  It is also
often assumed that the IMF is a standard power-law (or power-laws with
different indices in different mass ranges) with no primordial radial variation
in the cluster \citep[e.g.,
][]{1959ApJ...129..608S,1979ApJS...41..513M,2001MNRAS.322..231K}.   However,
star formation simulations indicate that neither of the above simplifications
may be valid.  For example, star formation takes place over a time-period and
the formation timescale depends on many different physical processes including
the Jeans mass of the protostellar clouds, radiative feedback and turbulence 
\citep[e.g., ][]{2006ApJ...641L.121T}.  Moreover, depending on these detailed physical
processes stars of different mass ranges may form over different timescales
\citep[e.g., ][]{2007MNRAS.381L..40D,2009ApJ...699..850K}.  In addition,
deviations from the standard MFs are observed in many clusters both at the high
mass and the low mass ends \citep[e.g., ][]{2004MNRAS.354..367E}.  In
particular at the high mass end observed MFs are generally top-heavy compared
to standard power-laws in dense cluster cores like the Arches cluster and
starburst regions
\citep{2002A&A...394..459S,2006ApJ...653L.113K,2009MNRAS.394.1529D}.
Mass segregation is also observed in old globular clusters \citep[e.g.,
][]{1993A&A...273..100S,1998ApJ...492..540H,2008ApJ...685..247B}.  
Whether the observed mass segregation is imprinted from the star formation 
epoch or it is attained via dynamical evolution of the cluster at a later time is still an 
open question.  
Although a conclusive general answer to this question does not exist, 
it has been shown in the case of the very young Trapezium cluster in the Orion 
nebula that the observed degree of mass segregation cannot be explained 
simply via dynamical mass segregation and a preferential formation of high-mass 
stars near the center is necessary to explain the observations 
\citep{1998MNRAS.295..691B}.     

In order to bypass the above mentioned assumptions a detailed model for the
star formation processes during the formation of the cluster is needed.
Interstellar clouds contract and fragment due to density perturbations to form
pre-stellar cores (PSC) that collapse on timescales that depend on their Jeans
masses to form stars \citep[e.g., ][]{1966PThPh..36..515N}.  Depending on the
compactness of the assortment of these PSCs they can merge with other PSCs and
grow before they collapse to form stars \citep[e.g.,
][]{1979ApJ...229..242S,2003MNRAS.338..817E,2007ApJ...661..262D}.  Using this
simple model \citet{2007MNRAS.381L..40D} have attempted to eliminate the
above-mentioned assumptions regarding cluster IMFs by building an IMF from 
a distribution of PSCs in a molecular cloud.  In their simplified model they assume that each PSC
collapses to form a single star of mass proportional to the parent PSC.
Comparing the collision timescale between the PSCs with the collapse timescale
of the PSCs they show that near the central regions of a dense cluster
the PSCs collide and grow in mass before they can collapse to form a star.
Thus, near the core where the number density of the PSCs is significantly
higher than that in the halo, it is more likely to form stars that are heavier
than stars formed from PSCs in the halo.  Using the Arches cluster as a
particular example they show that their semi-analytical model naturally results
in a mass segregated cluster similar to the Arches cluster and the observed 
top-heaviness near the center is thus explained as imprinted from the star 
formation epoch. 

The effects of primordial mass segregation on the global dynamical evolution of
clusters have been explored previously in a parametric way 
showing interesting differences in the global properties of those 
clusters and their dynamical evolution 
\citep[e.g.,][]{2008ApJ...685..247B,2004ApJ...604..632G}.  
The observed top-heavy MF in the central regions of the Arches cluster
\citep[e.g.,][]{2006ApJ...653L.113K,2007JKAS...40..153K} thus makes it an
interesting candidate for such studies.  If the observed radially dependent 
top-heaviness in Arches MF is indeed primordial, then this is a direct measurement 
of the degree of mass segregation at formation of this cluster, thus it eliminates 
the necessity for parameterization.  

The Arches cluster is also unique for the following reasons.  
The cluster has a 
projected distance of $\sim 30\,\rm{pc}$ from the Galactic center \citep{1995AJ....109.1676N,2002ApJ...565..265P,2006ApJ...653L.113K,2007JKAS...40..153K}.  Estimates from detailed proper 
motion observations with a baseline of $\sim 4\,\rm{yr}$ indicate that the real 
Galactocentric distance of the Arches cluster is $\approx 27\,\rm{pc}$ 
\citep{2008ApJ...675.1278S}.  Thus the Arches is the closest massive cluster from the 
Galactic center.  
The proximity of the 
cluster from the Galactic center necessitates this cluster to be compact to be saved 
in the face of possible early tidal disruption.  
\citet{1998Natur.394..448S} a decade ago found $\gtrsim 100$ O stars within $0.6\,\rm{pc}$ 
of the cluster, indicating Arches to be the densest young cluster in the Galaxy.  The 
high estimated total cluster mass ($\sim 10^4-10^5\,\rm{M_\odot}$), and the high luminosity 
$10^8\,\rm{L_\odot}$ thus make this young cluster a power house cluster for Galactic 
standards \citep[e.g.,][]{1998Natur.394..448S}.    
The central stellar density of 
the Arches cluster is $\sim 10^5\,\rm{M_\odot/pc^3}$ \citep[e.g.,][]{1998Natur.394..448S,2007MNRAS.378L..29P}, 
which is typical only for some old Galactic globular clusters.    The core radius ($r_c$) 
of the Arches cluster is $\lesssim 0.2\,\rm{pc}$ 
\citep[e.g.,][]{1999ApJ...525..750F,2002A&A...394..459S,2006JPhCS..54..217S}.  
The 
high central density and the high mass thus makes the Arches cluster a very good 
candidate to study for possibility of collisional runaway.  The Arches cluster is also 
very young, the cluster age ($t_{cl}$) is only $2\pm1\,\rm{Myr}$ 
\citep[e.g.,][]{2008ApJ...675.1278S}.  The timescale for a successful collisional runaway 
is only $\sim 3\,\rm{Myr}$ since past that age, the cluster is depleted of its high-mass 
stars due to Supernova explosions \citep[e.g.,][]{2006MNRAS.368..141F,2007ASPC..367..707F}.  
Hence, if conducive for a collisional runaway, the Arches cluster may be undergoing that 
process at the present time.        

Two main scenarios to create the observed top-heavy radially dependent MF have been
suggested in previous studies: 1. The observed MF is primordial, as mentioned
earlier \citep[e.g., ][]{2007MNRAS.381L..40D}.  2. The observed MF is a result
of dynamical evolution starting from a standard IMF with no primordial radial 
variation \citep[e.g.,][]{2007MNRAS.378L..29P}.  Using numerous numerical 
simulations we study the 
dynamical evolution of the Arches cluster assuming the observed MF was indeed 
primordial.  We compare the results from these simulations with the results from another 
set of simulations where the initial cluster properties are kept unchanged but 
a standard IMF with no radial variation is used instead.  We investigate whether it is 
possible to attain the 
observed degree of mass segregation starting from an unsegregated cluster.  
We verify that the MFs and cluster parameters of the simulated clusters at 
$t_{cl}\approx2\,\rm{Myr}$ are consistent with observations.  
We then investigate the differences in the dynamical evolution of the 
simulated Arches-like clusters with the two above-mentioned choices of IMFs focusing 
on the possibilities of a collisional runaway.  Furthermore, we study the final fate and 
discuss the possible observational signatures of a collisional 
runaway depending on the stellar wind prescription.  The arrangement 
of this paper is as follows.  
In
\S\ref{numerical} we describe our simulations.  In \S\ref{results} we summarize
our key results.  This is followed by \S\ref{effects} where we discuss the key observable 
signatures of a collisional runaway.  We summarize and conclude in \S\ref{conclusion}.      
\section{Initial Conditions} \label{numerical} 
We simulate clusters consisting initially of $65536$ stars with positions and
velocities chosen according to a King profile \citep{1966AJ.....71...64K} with dimensionless
central potential $W_0$ in the range from $5$ to $9$. The initial virial radius ($r_v$) of
the cluster is set to $0.5\,\rm{pc}$, which results in simulated clusters that have 
characteristic radii, e.g., $r_c$ and tidal radius ($r_t$) at $\approx 2\,\rm{Myr}$ 
close to the observed 
values for the Arches cluster (Table\ \ref{tab:runs}, see also \S\ref{intro} for observed values).  
Our simulated clusters are initially Roche lobe filling.  
Numerical simulations by \citet{2008MNRAS.389L..28G} show that 
dynamical evolution and tidal truncation makes initially underfilled and overfilled 
clusters Roche lobe filling and Roche lobe limited at timescales 
of the order of their half-mass relaxation time ($t_{r,h}$).  
Since the typical initial $t_{r,h}$ for our simulated Arches-like clusters is much 
lower ($\sim 10^3\,\rm{yr}$) than the current age of Arches, the above assumption 
is justified.  
         
All the simulated 
clusters are assumed to be in a circular orbit around the Galactic center.  Although 
the cluster orbit may be eccentric, 
the variations of
its distance to the Galactic center are not expected to be significant for this
study \citep{2008ApJ...675.1278S}.  
We simulate all clusters at a fixed Galacto-centric distance since given the present 
day mass and Galactic location of the Arches cluster, inspiral timescale from a larger 
distance is more than an order of magnitude longer than its estimated age 
\citep{2005ApJ...628..236G, 2008ApJ...675.1278S}.  
One may argue however, that Arches initially was a lot more massive 
\citep[needs to be $> 10^6\ \rm{M_\odot}$, ][]{2003ApJ...597..312K} and has lost most 
of its mass during its evolution and inspiral and is now observed at its current position 
just before complete disruption.  We find in our simulations that the total mass 
loss from escaping stars through the tidal boundary and from stellar evolution 
typically is only $\sim 0.04$ of the initial cluster mass during the first $2\,\rm{Myr}$.  
Since, at the current position of Arches the tidal effects of the Galaxy is the 
maximum if inspiral scenario is to be believed, this mass loss rate is an upper limit.  
Thus it is unlikely that Arches was initially more massive than $10^6\,\rm{M_\odot}$.  

\begin{figure}
\begin{center}
\plotone{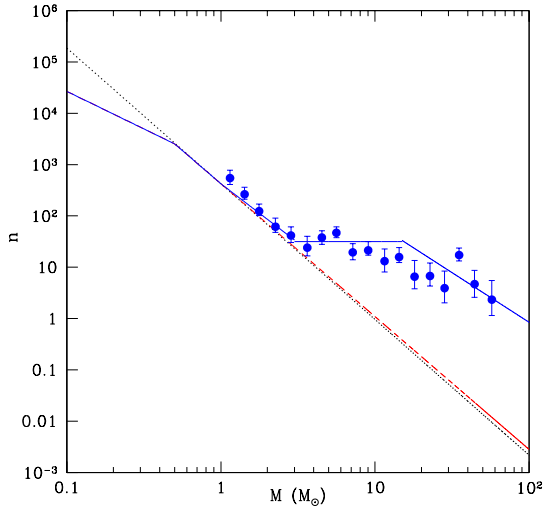}
\caption[IMFs]{Comparison of the {\tt Seg} IMF within $2r_c$ from the cluster center of 
mass with other standard IMFs and the observed data points.  Black dotted line is the 
standard Salpeter IMF, red dashed line is the \citet{2001MNRAS.322..231K} IMF, and 
The Blue dots and the error-bars are extracted using ADS's Dexter applet 
\citep{2001ASPC..238..321D} from \citet{2007MNRAS.381L..40D} showing the observed 
MF within $2r_c$ in Arches. The blue solid line is an approximation of the observed 
MF within $2r_c$ in the Arches cluster using broken 
power-laws of the form $n\propto M^{-\alpha}$, where $\alpha$ changes in different mass 
ranges.  The $\alpha$ values are given in the form $M_1(\alpha)M_2$ denoting the 
value of $\alpha$ in the mass range $M_1$ to $M_2$ in $\rm{M_\odot}$: $0.1(1.3)0.5(2.3)1.0(2.04)3.0(0.)15(1.72)100$.  
For $r>2r_c$, the IMF {\tt Seg} is standard Kroupa 2001.  
}
\label{plot:IMF}
\end{center}
\end{figure}

For each cluster profile we perform two sets of simulations using different
IMFs to assign masses to stars.  For the set of simulations with initially 
non-segregated clusters we use the IMF in
\citet{2001MNRAS.322..231K} for the whole cluster (henceforth, {\tt Nseg}).  
For the second set we choose a radially dependent IMF (henceforth, {\tt Seg}), 
with a central
top-heavy mass function for $r \leq 2r_c$, very similar to the observed one of 
\citet[][Figure\ \ref{plot:IMF}]{2006ApJ...653L.113K} and a standard \citet{2001MNRAS.322..231K} IMF outside.    
Initial cluster properties and key results are summarized in
Table\ \ref{tab:runs}.  In order to make sure that our results are robust
against statistical fluctuations we repeat each simulation $2$ times varying the
random seed for generating the positions and velocities for each cluster with the 
same IMF and initial $W_0$. In order to advance the clusters in time we use the direct $N$-body
code NBODY4 \citep{2003gnbs.book.....A} which includes the tidal effects of the
Galaxy, the evolution of stars using the integrated SSE/BSE package
\citep{2000MNRAS.315..543H,2002MNRAS.329..897H}, and makes use of the GRAPE6 hardware to speed-up
the computation of the gravitational forces between stars. The evolution of each 
cluster is followed for $5\,\rm{Myr}$, which is longer than the approximate 
time when the first supernova explosions occur, turning the most massive stars into 
black holes and consequently ending any possible runaway-growth
\citep{2002A&A...394..345F,2006MNRAS.368..121F,2006MNRAS.368..141F}.  We do not 
have primordial binaries in our simulated clusters, although binaries created via 
three-body interactions are automatically included.  
The initial $\sim 10\,\rm{Myr}$ evolution of a star 
cluster is dominated by stellar evolution mass loss and two-body relaxation.  Strong 
interaction mediated by binaries become a more important factor only at later times so 
far as the global evolution of the cluster is concerned.  Including binaries, however, may 
decrease the mass segregation timescale and increase collision rate to some degree.  
Nevertheless, since the binary fraction in Arches is not known observationally, we choose 
not to consider primordial binaries rather than making a wild guess; the additional 
benefit is a significant reduction in computational cost.  
\begin{deluxetable*}{ccccccccccccc}
\tabletypesize{\footnotesize}
\tablecolumns{9}
\tablewidth{0pt}
\tablecaption{List of simulations \label{tab:runs}}
\tablehead{
\colhead{Name} & 
\multicolumn{2}{c}{$M\ (10^4\ \rm{M_\odot}$)\tablenotemark{a}} & 
\colhead{King} & 
\colhead{IMF\tablenotemark{c}} & 
\multicolumn{4}{c}{$r\,(\rm{pc})$\tablenotemark{d}} & 
$M_{\rm{max}}$\,($\rm{M_\odot}$)\tablenotemark{e} & 
\#Coll\\
\colhead{} & 
\colhead{$0\,\rm{Myr}$} & \colhead{$2\,\rm{Myr}$} &
\colhead{$W_0$ \tablenotemark{b}} & 
\colhead{ } &
\colhead{$r_c$} & \colhead{$r_t$} & \colhead{$r_v$} & \colhead{$r_{hl}$} &
\colhead{} & 
\colhead{} 
}
\startdata
{\tt SW9-1} & $6.2$ & $5.9$ & $9$ &  & $0.17\pm0.05$\tablenotemark{f} & $19.7\pm0.7$ & $0.73\pm0.03$ & $1.14\pm0.06$ & $327$, $252$ & $19$ \\
{\tt SW9-2} & $6.3$ & $6.1$ & $9$ &  & $0.12\pm0.04$ & $17.3\pm0.6$ & $0.66\pm0.01$ & $1.17\pm0.0.07$ & $356$, -\tablenotemark{g} & $9$ \\
{\tt SW8-1} & $9.2$ & $8.8$ & $8$ &  & $0.11\pm0.03$ & $15.2\pm0.7$ & $0.62\pm0.01$ & $0.74\pm0.04$ & $806$, $326$ & $30$ \\
{\tt SW8-2} & $9.0$ & $8.6$ & $8$ &  & $0.07\pm0.02$ & $9.4\pm0.5$ & $0.48\pm0.01$ & $0.36\pm0.02$ & $1097$, $647$ & $47$ \\
{\tt SW7-1} & $9.2$ & $8.9$ & $7$ & {\tt Seg} & $0.10\pm0.02$ & $9.6\pm0.3$ & $0.49\pm0.01$ & $0.46\pm0.02$ & $683$, $278$ & $22$ \\
{\tt SW7-2} & $10.0$ & $9.7$ & $7$ &  & $0.13\pm0.04$ & $10.8\pm0.3$ & $0.52\pm0.01$ & $0.55\pm0.02$ & $703$, $428$ & $15$ \\
{\tt SW6-1} & $12.8$ & $12.4$ & $6$ &  & $0.10\pm0.02$ & $8.4\pm0.2$ & $0.46\pm0.01$ & $0.40\pm0.02$ & $140$, $117$ & $13$ \\
{\tt SW6-2} & $12.9$ & $12.5$ & $6$ &  & $0.17\pm0.01$ & $8.7\pm0.2$ & $0.47\pm0.01$ & $0.44\pm0.01$ & $382$, $144$ & $26$ \\
{\tt SW5-1} & $15.9$ & $15.3$ & $5$ &  & $0.19\pm0.01$ & $7.6\pm0.2$ & $0.43\pm0.01$ & $0.40\pm0.01$ & $126$, $125$ & $14$ \\
{\tt SW5-2} & $15.5$ & $15.0$ & $5$ &  & $0.20\pm0.01$ & $7.4\pm0.2$ & $0.43\pm0.01$ & $0.39\pm0.01$ & $156$, $107$ & $13$ \\
\hline \\
{\tt NSW9-1} & $3.23$ & $3.18$ & $9$ &  & $0.04\pm0.01$ & $12.0\pm0.3$ & $0.55\pm0.01$ & $0.8\pm0.2$ & $339$, - & $14$ \\
{\tt NSW9-2} & $3.20$ & $3.18$ & $9$ &  & $0.04\pm0.04$ & $11.9\pm0.2$ & $0.55\pm0.01$ & $0.2\pm0.2$ & $238$, $104$ & $15$ \\
{\tt NSW8-1} & $4.2$ & $4.1$ & $8$ &  & $0.18\pm0.09$ & $12.2\pm0.8$ & $0.55\pm0.02$ & $0.57\pm0.06$ & $136$, $131$ & $14$ \\
{\tt NSW8-2} & $3.2$ & $3.2$ & $8$ &  & $0.06\pm0.02$ & $8.9\pm0.2$ & $0.47\pm0.01$ & $0.28\pm0.06$ & $135$, - & $8$ \\
{\tt NSW7-1} & $3.15$ & $3.14$ & $7$ & {\tt NSeg} & $0.08\pm0.04$ & $7.6\pm0.1$ & $0.4\pm0.1$ & $0.4\pm0.1$ & -, - & $8$ \\
{\tt NSW7-2} & $3.23$ & $3.21$ & $7$ &  & $0.11\pm0.03$ & $7.6\pm0.3$ & $0.436\pm0.008$ & $0.24\pm0.05$ & $130$, - & $5$ \\
{\tt NSW6-1} & $3.3$ & $3.2$ & $6$ &  & $0.08\pm0.02$ & $7.3\pm0.1$ & $0.426\pm0.004$ & $0.26\pm0.02$ & $116$, - & $2$ \\
{\tt NSW6-2} & $3.2$ & $3.2$ & $6$ &  & $0.11\pm0.02$ & $6.8\pm0.1$ & $0.412\pm0.003$ & $0.37\pm0.04$ & -, - & $3$ \\
{\tt NSW5-1} & $3.2$ & $3.1$ & $5$ &  & $0.15\pm0.05$ & $15.3\pm0.5$ & $0.62\pm0.01$ & $0.93\pm0.07$ & $233$, $112$ & $16$ \\
{\tt NSW5-2} & $3.2$ & $3.2$ & $5$ &  & $0.08\pm0.03$ & $6.9\pm0.1$ & $0.42\pm0.01$ & $0.12\pm0.02$ & $147$, $104$ & $7$ \\
\enddata
\tablenotetext{a}{Bound mass at the age ($t_{cl}$) of the simulated cluster.  }
\tablenotetext{b}{King profile \citep{1966AJ.....71...64K}  }
\tablenotetext{c}{The range in mass is $[0.1-100]\,\rm{M_\odot}$}
\tablenotetext{d}{at $t_{cl} = 2\,\rm{Myr}$  } 
\tablenotetext{e}{Highest mass reached for the two most massive stars in the simulation.  }
\tablenotetext{f}{All quoted errors are $1\sigma$ error-bars from statistical fluctuations between the snapshots over which the cluster properties are averaged for the the simulated clusters.  }
\tablenotetext{g}{Mass does not grow above $100\ \rm{M_\odot}$ within $4\,\rm{Myr}$ }
\end{deluxetable*}
\section{Results}
\label{results}
In this section we present the results from our numerical simulations.  We show in detail 
the results for clusters with $W_0=8$ and we summarize more briefly the results for all 
other runs.  
\subsection{Evolution of the Mass Function and Observational Constraints}
\label{MF_comp}
Here we show, how well the initially segregated ({\tt Seg}) and unsegregated ({\tt NSeg})
cluster models can reproduce the observed level of mass segregation at the current 
age ($2\,\rm{Myr}$)
of the Arches cluster.
As a condition for consistency, the simulated clusters should have a present day 
MF consistent 
with the observed MF at the age of the 
Arches cluster.  We compare the MF from the simulated clusters with ({\tt Seg}) and 
without ({\tt NSeg}) mass segregation at $t_{cl} = 2\,\rm{Myr}$ with 
the observed MF.  
\begin{figure}
\begin{center}
\plotone{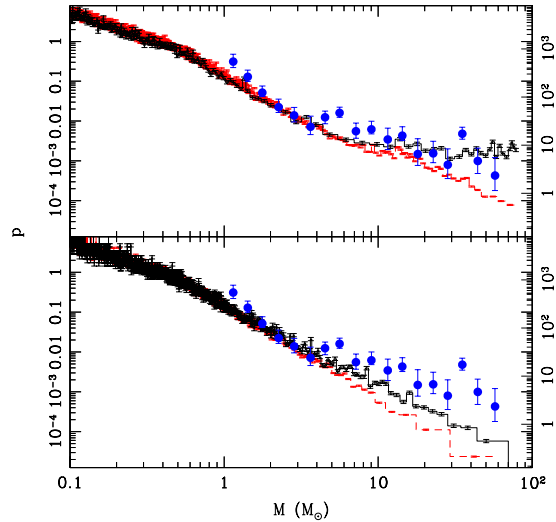}
\caption[PDF MF]{MFs of the simulated clusters for 
runs {\tt SW8-1} (top panel) and {\tt NSW8-1} (bottom panel) 
at $2.0\,\rm{Myr}$.  The 
red (dashed) and the black (solid) histograms show MFs outside and inside 
$r = 2r_c$, respectively.  The blue dots show the present day observed MF, taken 
from \citet{2007MNRAS.381L..40D} using the Dexter application from ADS.  The 
simulated histograms are normalized so that the area enclosed is unity.  The 
observed points have the same normalization as used by \citet[][see right labels on 
the vertical axis]{2007MNRAS.381L..40D}.  Both simulations show some degree of mass 
segregation.  
The primordially unsegregated cluster {\tt NSW8-1} do not 
attain the observed degree of mass-segregation at $t_{cl}=2\,\rm{Myr}$.  In contrast, 
the simulated MF matches the observed MF reasonably well for run {\tt SW8-1} with 
the primordially mass-segregated cluster with the {\tt Seg} IMF.     
}
\label{plot:MF}
\end{center}
\end{figure} 
\begin{figure}
\begin{center}
\plotone{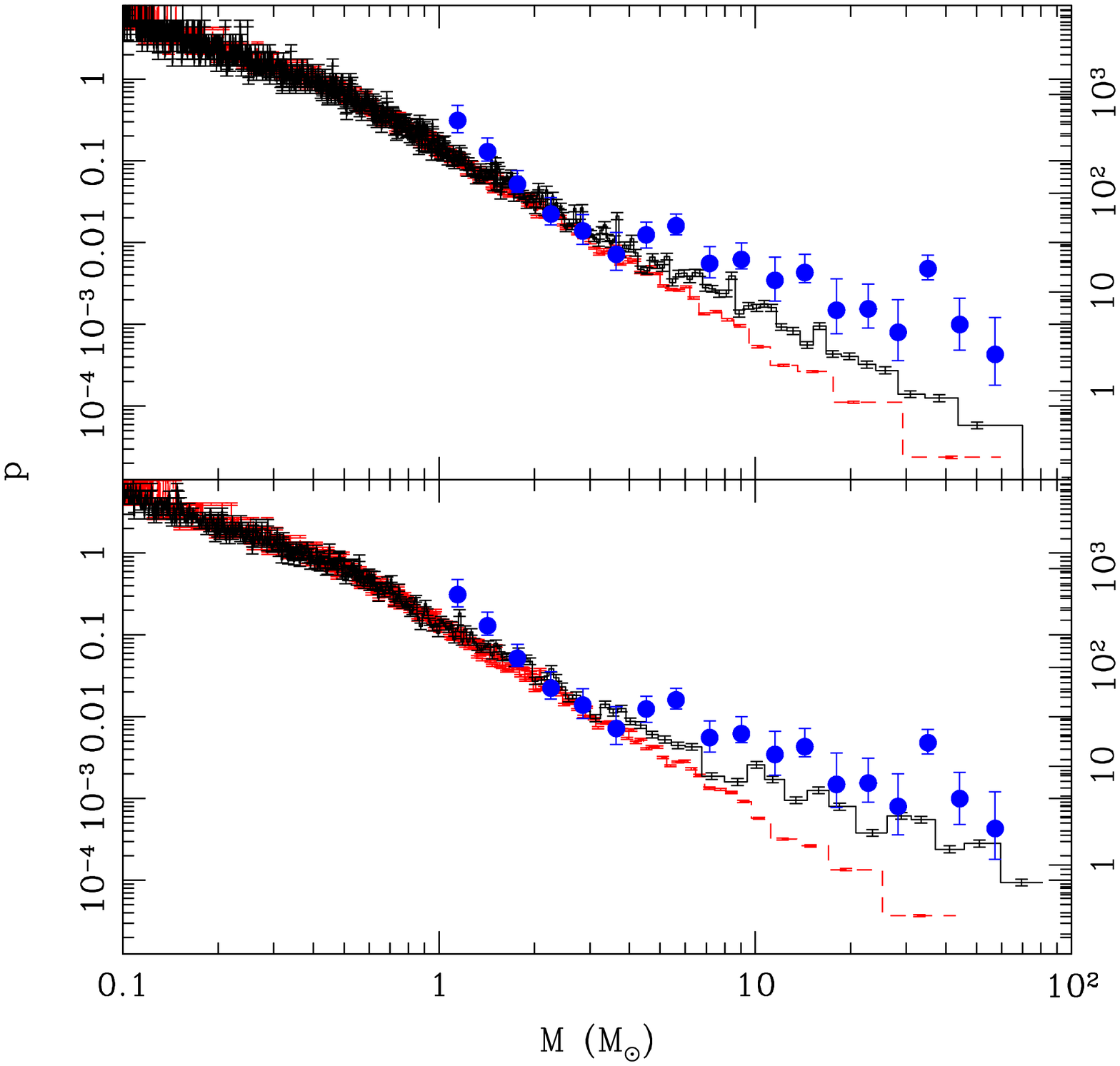}
\caption[PDF MF: SE]{MFs of the simulated clusters for 
runs {\tt NSW8-1} with (top panel) and without (bottom panel) stellar evolution 
at $2.0\,\rm{Myr}$.  The 
red (dashed) and the black (solid) histograms show MFs outside and inside 
$r = 2r_c$, respectively.  The blue dots show the present day observed MF, taken 
from \citep{2007MNRAS.381L..40D} using the Dexter application from ADS.  The 
simulated histograms are normalized so that the area enclosed is unity.  The 
observed points have the same normalization as used by \citet[][right labels on 
the y axis]{2007MNRAS.381L..40D}.  In both cases the initial clusters are identical.  
The simulation including stellar evolution shows significantly lower degree of mass 
segregation than that without stellar evolution.  
}
\label{plot:MF-senose}
\end{center}
\end{figure} 
Figure\ \ref{plot:MF} shows the MF within $r = 2r_c$ for runs {\tt SW8-1} and 
{\tt NSW8-1} at $t_{cl} = 2\,\rm{Myr}$.  
Each mass bin contains $50$ stars so that 
the Poisson error is equal in each bin, and the MF is 
normalized such that its integral is unity.  
As it can be seen, while the evolved MF in the inner region of the {\tt Seg}
cluster is in good agreement with observations, this is not the case for the
{\tt NSeg} cluster.  For the {\tt NSeg} cluster the slope at the high mass end 
increases only marginally (from $-2.3$ to $-1.6$).  While for the {\tt Seg} cluster 
the MF becomes flatter, but remains still compatible with observations.  Same 
results are obtained when the MFs are compared at $t_{cl} \approx 3\,\rm{Myr}$.  
This 
indicates that the degree of mass segregation observed in the Arches cluster is not 
likely to result simply from the cluster's dynamical evolution during its 
lifetime without any primordial mass segregation.  It also indicates that, the MF 
at $t_{cl} = 2\,\rm{Myr}$ for the simulated cluster is consistent if the 
observed MF of the Arches cluster truly is imprinted from its star formation epoch 
as suggested by \citet{2007JKAS...40..157D}.       

We should note that for the {\tt NSeg} IMF, a much higher degree of mass segregation 
can be achieved if stellar evolution is not included in the simulations.  
We find that if we turn off stellar evolution, the degree of flattening in the MF 
of the simulated cluster at $\sim 2\,\rm{Myr}$ increases 
significantly compared to that with stellar evolution (Figure\ \ref{plot:MF-senose}).  
Although the wind mass 
loss from the high-mass stars due to stellar evolution is negligible ($<4\%$) compared 
to the total cluster mass in these simulations during the 
first $2\,\rm{Myr}$, this mass 
is lost from the most massive stars in the cluster, which in turn reside near the center due 
to mass segregation.  Thus this mass is lost from the deepest part of the cluster 
potential and can have a perceptible effect on the global properties of the cluster.    
From our results we conclude that the 
mass loss via stellar evolution from the massive stars can inhibit mass segregation to 
a certain extent and simulations done without taking that physical effect into account will 
overestimate the amount of mass segregation.  Since this effect is only predominant for the 
most massive stars at a given time in the cluster, the quantitative effect will 
depend on the number of high-mass stars present in the cluster.  
Despite a lower total number of initial stars ($N \approx 12K$ and $24K$), simulations in 
\citet{2007MNRAS.378L..29P} had more massive stars than our {\tt NSeg} clusters 
because of the truncated IMF they used.  Thus, if stellar evolution is included, mass 
segregation should have been even more strongly suppressed in their simulations compared to 
our simulations with the {\tt NSeg} clusters (Figure\ \ref{plot:MF-senose}).  This 
further supports the notion that the level of observed mass segregation in the Arches 
cluster cannot have a fully dynamical origin.   
\subsection{Choice of IMF and Implications for Collisional Runaway}
\label{runaway}
In general we find that for the clusters with primordial mass-segregation 
(set {\tt Seg}) the number of 
collisions is always significantly larger compared to the clusters with a 
standard Kroupa IMF without any primordial mass segregation (set {\tt NSeg}; see 
Table\ \ref{tab:runs}), for example, in the case 
$W_0=8$ by more than a factor of two ($30$ for 
run {\tt SW8-1} and $14$ for run {\tt NSW8-1}) within the first $4\,\rm{Myr}$. 
This is mainly due to the 
top-heavyness of the Arches IMF within $r \leq 2r_c$ (\S\ref{numerical}) and 
the larger overall mass of our clusters with primordial mass segregation, 
resulting in a larger number of massive stars with large collision cross 
sections. Consequently, the runaway growth of a massive star is much more 
likely to take place in clusters from set {\tt Seg}.  

\begin{figure}
\begin{center}
\plotone{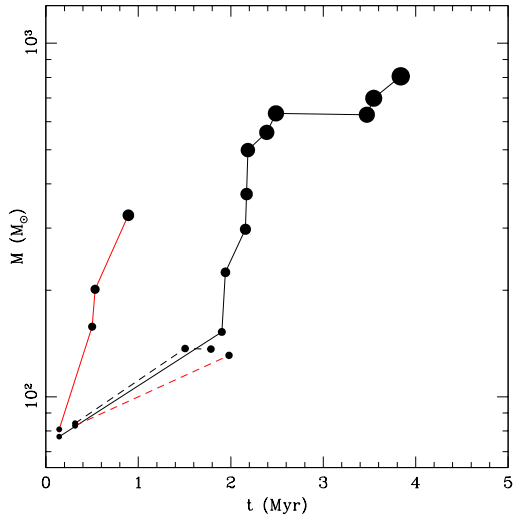}
\caption[collisional growth of mass]{The growth of mass in time for the two most 
massive stars in the cluster are plotted for runs {\tt SW8-1} (solid lines) 
and {\tt NSW8-1} (dashed lines).  Each dot denotes the time when the 
particular star changed mass through collision.  The area of each dot is 
proportional to the stellar mass.  For the {\tt Seg} IMF a collisional runaway is 
observed starting around $2\,\rm{Myr}$.  For the {\tt NSeg} IMF using the same initial 
conditions there is no runaway.  }
\label{plot:tvsm_w8-1}
\end{center}
\end{figure}
\begin{figure}
\begin{center}
\plotone{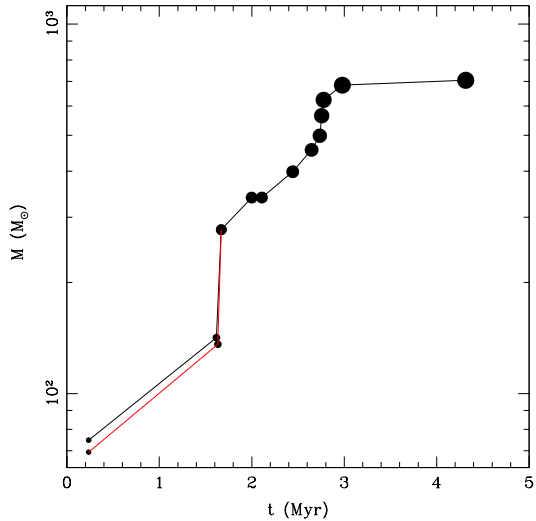}
\caption[collisional growth of mass]{Same as Figure\ \ref{plot:tvsm_w8-1} but for 
runs {\tt SW7-1} and {\tt NSW7-1}.  No star in run {\tt NSW7-1} grows to $>100\,\rm{M_\odot}$ 
within the first $4\,\rm{Myr}$, hence are not not shown on this plot.  }
\label{plot:tvsm_w7-1}
\end{center}
\end{figure}

 
As an example, Figure\ \ref{plot:tvsm_w8-1} shows the mass of the two most massive stars over time for the 
cluster with the {\tt Seg} IMF (run {\tt SW8-1}) and the {\tt NSeg} IMF 
(run {\tt NSW8-1}). In the 
{\tt Seg} case there is a clear evidence of a collisional runaway producing 
two very massive stars. The most massive star in this case attains a mass 
of $800\,\rm{M_\odot}$ through $12$ successive collisions. The second most massive 
star also grows significantly and reaches $300\,\rm{M_\odot}$ through $4$ successive collisions.  
A clear sign of a collisional runaway is a steep increase of the stellar mass through 
successive collisions.  Figure\ \ref{plot:tvsm_w8-1} clearly shows that at $t_{cl} \simeq 2\,\rm{Myr}$ 
the most massive star enters a runaway regime.  
The whole process of the runaway takes place 
between $\sim 2$--$4\,\rm{Myr}$. In case of {\tt NSeg} IMF the two most 
massive stars experience only two to three collisions and there is no runaway 
happening. We obtain qualitatively similar results when changing the random 
seed to generate other equivalent initial conditions. For instance, in the run {\tt SW8-2} 
with {\tt Seg} IMF we see a runaway, where the most massive star grows to 
$\sim 10^3\,\rm{M_\odot}$ within the first $4\,\rm{Myr}$ starting from 
about $90\,\rm{M_\odot}$, through $17$ successive collisions. The corresponding 
run with the {\tt NSeg} IMF, {\tt NSW8-2}, did not exhibit any runaway growth 
(Table\ \ref{tab:runs}).  A similar result is obtained for simulated clusters with other 
$W_0$ values.  For example, Figure\ \ref{plot:tvsm_w7-1} shows the growth of the two most 
massive stars in the simulated clusters for runs {\tt SW7-1} and {\tt NSW7-1}, both clusters 
with $W_0 = 7$, but the first from the {\tt Seg} and the latter from the {\tt NSeg} set.  Again, 
the run from set {\tt Seg} shows clear evidence of a collisional runaway, whereas 
the equivalent cluster from set {\tt NSeg} shows none. (For a 
full list of simulation results with other initial $W_0$ values, see Table\ \ref{tab:runs}.) 
Note that, in the collisional runaway of the stars in our simulations, 
we find that the stars can grow even after $3\,\rm{Myr}$ via collisions.  
This is due to the collisional rejuvenation prescription in BSE, which assumes complete 
mixing \citep{2000MNRAS.315..543H}.  Thus the collisional stars live longer compared 
to their normal lifetime.  This assumption is of course a simplification 
and the amount of mixing and thus rejuvenation depends 
on the details of the kinematics of the collisions 
\citep[e.g., ][]{1997ApJ...487..290S,2008A&A...488.1017G} .  
However, this assumption gives an upper limit of the degree of rejuvenation and 
lets the stars grow for a longer time than with 
a more realistic rejuvenation prescription.  In this study for the considerations of a 
collisional runaway we take collisional sequences only upto $t_{cl} = 4\,\rm{Myr}$.  

\begin{figure}
\begin{center}
\plotone{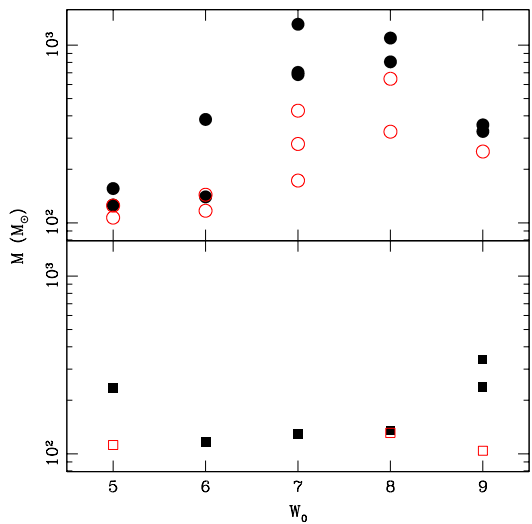}
\caption[$W_0$ and runaway]{Masses of the two most massive stars are shown for 
each simulated cluster with the {\tt Seg} (top panel) and {\tt NSeg} (bottom panel) 
IMFs as a function of the initial King concentration parameter $W_0$.  The 
filled point and the open point show the highest and the second highest mass attained 
during the first $4\,\rm{Myr}$ of evolution.  Cases where 
the mass is not above $100\ \rm{M_\odot}$ are not shown since the initial high mass cut-off of 
the IMF is $100\ \rm{M_\odot}$ in our simulations.  Multiple realizations of the same cluster 
in each case is simulated to ascertain robustness against statistical fluctuations.  Note that 
for the {\tt Seg} IMF with $W_0$ $7$ and $8$, there is clear indication of double runaways.  }
\label{plot:wvsm}
\end{center}
\end{figure}

A very interesting aspect of these results with the {\tt Seg} 
IMF is that, many of these runs lead to a double collisional runaway, 
one being stronger than the other (e.g., runs {\tt SW8-1,2}, {\tt SW7-1,2}; see 
Table\ \ref{tab:runs} for a full list) creating two very massive stars (VMS).  
Double runaways have been studied and observed 
in some previous simulations 
with primordial binaries \citep[e.g., ][]{2006ApJ...640L..39G}, however, a double runaway 
even without primordial binaries has not been observed before.  Due to the flat mass spectrum 
in the high mass end of the {\tt Seg} IMF, there is a larger reservoir of high-mass stars 
increasing the chance of a double runaway.  A double runaway does not 
necessarily mean black hole binaries, as discussed in \citet{2006ApJ...640L..39G}, since 
the two VMSs may still collide with each other at a later time but before compact object 
formation (e.g., {\tt SW8-2}, {\tt SW7-1}).  The signature of a double runaway and production 
of two VMSs seems to be common in all of our simulations with the {\tt Seg} IMF, 
where the initial concentration is conducive to a collisional runaway, in particular for runs 
with King concentration parameters $W_0 = 7, 8, 9$ (Figure\ \ref{plot:wvsm}).  

Similar results are obtained for clusters with other initial concentrations or 
$W_0$ values. The main difference is in the final mass of the most massive 
stars, shown in Figure\ \ref{plot:wvsm}.  The final 
stellar mass is also subject to statistical fluctuations to a certain extent.  
The collisional runaway with the {\tt Seg} 
IMF and $W_0 = 6$ is weak, whereas, for $W_0 = 5$ there is no runaway 
(Table\ \ref{tab:runs}).  For each of these runs with the {\tt Seg} IMF, the corresponding 
simulations with the same initial $W_0$ but the {\tt NSeg} IMF do not exhibit any runaway 
and do not produce the VMSs.  
However, for $W_0 = 9$, both the {\tt Seg} and the {\tt NSeg} IMF simulations show runaways 
and the masses of the collisionally created most massive stars are comparable.   
In the {\tt Seg} case $W_0 = 9$ shows a weaker 
collisional runaway in our simulated examples.  
This is because from a King model with 
$W_0 = 8$ to one with $W_0=9$, the core radius decreases considerably, thus the 
top-heavy IMF is within a smaller volume.  As a result the total initial mass as well 
as the number of high-mass stars decrease.  From our simulations, we find that with 
the {\tt Seg} IMF, $W_0 = 7$ and $8$ are optimal for strong collisional runaways.  On 
one hand a lower $W_0$ decreases the probability of a runaway because of the lower 
central density.  On the other hand, a higher $W_0$ gives the same result although the 
central density is higher, because of the lower overall mass of the cluster as well as 
the lower number of high-mass stars.      

\subsection{Enhanced Mass Loss via Stellar Winds}
\label{results_winds}
\begin{figure}
\begin{center}
\plotone{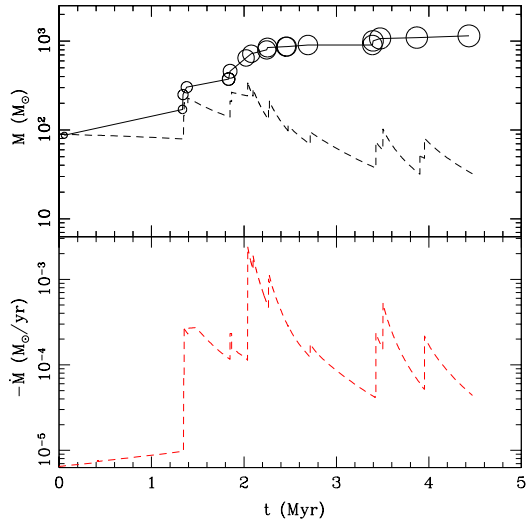}
\caption[mass loss]{Effect of wind mass loss for the run {\tt SW8-2}.  
Top panel: Solid line shows the mass growth 
via collision with the standard stellar evolution prescription in Nbody4.  The circles 
show the time when the collisions happened.  
The dashed line shows the same collision chain, only with the high wind mass loss 
prescription \citep{2009A&A...497..255G}.  Following each collision, the collision product 
is evolved and wind mass loss is taken into account.  Bottom panel: The wind mass loss 
rate as a function of time.  }
\label{plot:wind}
\end{center}
\end{figure} 
High-mass stars can lose a significant fraction of their mass through stellar winds 
\citep[e.g.,][]{1979ARA&A..17..275C,1986ARA&A..24..329C,2000ARA&A..38..613K}.  
\citet{2009A&A...497..255G} show that a VMS generated via stellar collisions 
may lose mass through stellar winds at such a high rate that its collisional growth 
may be quenched.  We explore this possibility, following their work, for 
our simulated clusters {\tt SW8-1} and {\tt SW8-2}, where collisional runaways have 
been observed.  
Following \citet{2009A&A...497..255G} each collision is treated using the ``Make Me 
a Massive Star" (MMAMS) software package \citep{2008MNRAS.383L...5G,2002ApJ...568..939L}, 
which gives the properties of the collision product.  This output is then imported into 
the Eggleton stellar evolution code \citep{1971MNRAS.151..351E,1972MNRAS.156..361E,1995MNRAS.274..964P} using the method of \citet{2008A&A...488.1007G}.  This method is repeated 
for every collision in the collision chain 
\citep[for a more detailed description see][especially Section\ $2.1$]{2009A&A...497..255G}.            

\begin{figure}
\begin{center}
\plotone{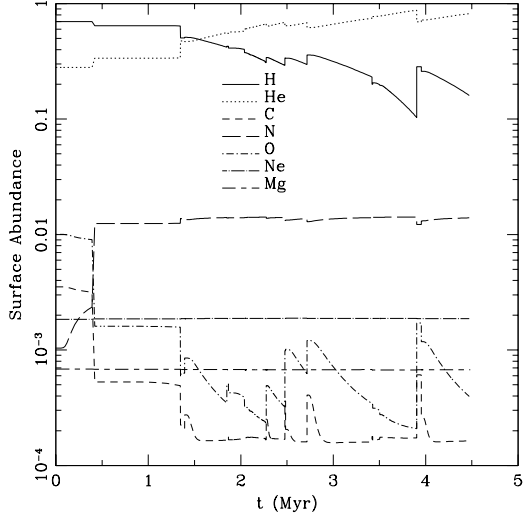}
\caption[abundance]{Time evolution of the surface abundance of the most massive 
star in the run {\tt SW8-2}.  Surface abundance of Hydrogen (H), Helium (He), Carbon (C), 
Nitrogen (N), Oxygen (O), Neon (Ne), and Magnesium (Mg) are shown.  Note the very 
high surface abundance of Helium.  }
\label{plot:abundance}
\end{center}
\end{figure}

Figure\ \ref{plot:wind} 
shows the time evolution of the most massive star for the run {\tt SW8-2} as an example.  
The solid line in the top panel shows 
the collisional growth of the stellar mass in the runaway observed in our run {\tt SW8-2}, 
using the standard wind prescriptions from BSE/SSE, included in NBODY4.  
The dashed line in the top panel tracks the mass of the star assuming the wind mass 
loss prescriptions suggested by \citet{2009A&A...497..255G}.  The bottom panel shows the 
mass loss rate via stellar winds as a function of time according to \citet{2009A&A...497..255G}.  
With this wind mass loss prescription, while collisions 
increase the stellar mass, the increased stellar mass in turn leads to 
enhanced wind mass loss.  
The growth of the stellar mass through the chain 
of collisions is quenched and instead of the $\sim 10^3\,\rm{M_\odot}$ VMS a star of 
only $\sim 30\,\rm{M_\odot}$ is produced with this wind prescription.  
At the final stage the star is burning Helium in 
its core and has a high surface Helium abundance ($X_{He,S} \gtrsim 0.8$)
(Figure\ \ref{plot:abundance}).  
We should mention that the modeling of stellar winds from high-mass stars is quite uncertain, 
and various treatments often come from extrapolations from observed stars at a lower 
mass.  Depending on the prescription wind mass loss rates may vary by orders 
of magnitude.  
If indeed the high wind mass loss prescription is correct, then a 
VMS will not be produced even when a collisional 
runaway takes place.  Interestingly, the mass lost via stellar winds due to the collisional growth 
of the stars will be extremely Helium rich which may provide observable signatures 
(more on this in \S\ref{wind}).  Similar results are obtained from the collision 
chain in run {\tt SW8-1}.    

\section{Possible observational Signatures of a Collisional Runaway}
\label{effects}
Here we discuss the possible observable signatures of a 
collisional runaway in a cluster.  In the previous section we have shown that depending 
on the stellar wind prescription a collisional runaway may create a VMS or the mass growth 
can be quenched through high wind mass loss.  
Here we consider both cases and discuss possible observable signatures of both 
possibilities.  

\subsection{Possible Creation of another Pistol Star}
\label{pistol}
\begin{figure}
\begin{center}
\plotone{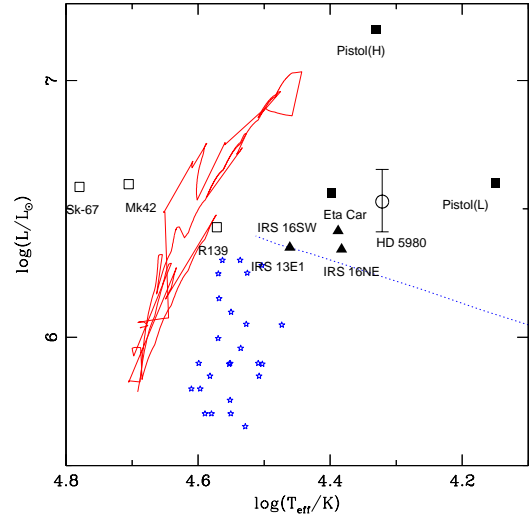}
\caption[T vs L]{The evolution of the most massive star created in run {\tt SW8-2} 
in the luminosity ($L$) vs effective temperature ($T_{eff}$) plane.  Solid red line shows the evolution of the most massive star 
for the enhanced wind mass loss prescription.  
The dotted line (blue) shows the Humphreys-Davidson limit \citep{1979ApJ...232..409H}.  
The observed luminous stars are also plotted for reference.  The filled black squares 
denote LBVs, open squares denote observed OB stars in the Magellanic clouds, 
filled triangles denote stars in GCs \citep[][and the references therein]{1998ApJ...506..384F}.  
Two points for the 
Pistol star denote high (H) and low (L) luminosity models \citep{1998ApJ...506..384F}.  The 
blue pentagons show the observed bright stars in Arches 
\citep[][and the references therein]{2009arXiv0909.3818C}.  }
\label{plot:lt}
\end{center}
\end{figure} 
The Quintuplet cluster is very similar to the Arches cluster.  It is also very near the 
Galactic center \citep[projected distance $\lesssim40\,\rm{pc}$; e.g.,][]{1990ApJ...351...83N,2002ApJ...565..265P}.  The Arches and the Quintuplet are also similar in their metallicities, 
and mass \citep{2009ApJ...691.1816N}.  The Arches is younger ($2\pm1\,\rm{Myr}$) than the Quintuplet  cluster ($\sim 4\,\rm{Myr}$).  The Pistol star is one of the most interesting and well 
studied objects in the Quintuplet.  This is one of the brightest stars ever observed 
\citep[Figure\ \ref{plot:lt};][]{1994MNRAS.268..194M,1996ApJ...461..750C,1997ApJ...474..275L,1998ApJ...506..384F}.  The inferred 
initial mass of the Pistol star is $200 - 250\,\rm{M_\odot}$.  The Pistol star is classified as 
a luminous blue variable \citep[LBV][]{1998ApJ...506..384F} or a B[e] 
\citep{1996ApJ...470..597M} star.  The stellar surface is extremely Helium rich and it has lost 
a significant fraction of its mass via Helium rich massive stellar winds 
\citep[e.g.][]{1995ApJ...447L..29F,1996ApJ...461..750C,1996ApJ...466..242T,1997ApJ...474..275L,1998ApJ...506..384F}.  From our results we find that the massive stars generated via collisional 
runaways may be a potential channel to produce a massive and Helium rich Pistol like star 
(Figure\ \ref{plot:wind}, \ref{plot:abundance}).  Figure\ \ref{plot:lt} shows the evolution of the most massive star 
found in run {\tt SW8-2} for the enhanced wind mass loss case as an example.  
The massive star produced through the collisional runaway spends a significant 
period of time (e.g., $1.4 - 2.3\,\rm{Myr}$ in case of the run {\tt SW8-2}) above the 
Humphreys-Davidson limit \citep{1979ApJ...232..409H} like 
the Pistol star.  Thus if the 
Arches was indeed created with a top-heavy IMF at the center as is observed today, then a 
Pistol-like star could be created at some point.  However, due to the uncertainty in 
the age estimate of the Arches cluster and the onset of the collisional runaway in simulations 
depending on the initial conditions it is not possible to predict whether a similar LBV 
star should already have been created in the Arches cluster and could be observed today.          

\subsection{IMBH Progenitors}
\label{IMBH}
The presence of supermassive black holes at the center of most galaxies is now well established 
\citep[e.g.,][]{1998Natur.395A..14R,2001AIPC..586..363K,2005ApJ...620..744G}.  Extrapolating 
the relation between the host mass and the black hole mass it has been postulated that massive 
clusters may also host black holes in the mass range $10^2$ -- $10^4\,\rm{M_\odot}$ 
\citep[e.g., see reviews by][]{2004cbhg.symp...37V,2004cbhg.symp..138R}.  These black 
holes are much less massive than the supermassive black holes at the centers of most galaxies 
but more massive than the stellar-mass black holes expected as remnants from high-mass 
stars.  Although no conclusive observational evidence for an IMBH has been obtained yet, 
there are several interesting candidates \citep[e.g.,][]{2009Natur.460...73F,2009arXiv0908.1115I}.  
The main proposed channel of formation for these IMBHs 
is through collisional runaways in dense clusters \citep[e.g.,][]{2002A&A...394..345F,2002ApJ...576..899P,2004ApJ...604..632G,2006ApJ...640L..39G,2006MNRAS.368..121F,2006MNRAS.368..141F,2007ASPC..367..707F,2006MNRAS.368..141F,2006MNRAS.368..121F}.  It is thus exciting to 
find young star clusters where a collisional runaway may have taken place.     

The simulated clusters with the {\tt Seg} IMF, in particular with $W_0 = 7$, $8$, and $9$ 
create VMSs via collisional runaway.  The counterparts of the same clusters with the more 
standard {\tt NSeg} IMF do not produce these VMSs.  If the 
winds do not quench the mass growth 
via collisions, then the VMSs can potentially be IMBH progenitors 
\citep[e.g.][]{1999ApJ...522..413F,2001ApJ...554..548F,2002ApJ...576..899P,2005ASPC..332..339H,2009arXiv0904.2784B}.  If an IMBH is indeed 
produced, there can be many observational signatures.  Even without primordial 
binaries, at the high densities attained at the core binaries can form dynamically from 
three-body interactions.  IMBHs being a lot more massive than the typical stars in its 
vicinity inevitably exchange into a binary or a higher order bound system 
\citep{2008ApJ...686..303G}.  
A stable mass transfer to the IMBH from its binary companion can be visible as 
ultra luminous X-ray sources.  
Inspiral and capture of compact objects such as stellar mass black holes and neutron 
stars into the IMBH can be a strong gravity wave source detectable 
with the Laser Interferometric Space Antennae (LISA).  The simulated clusters 
with the {\tt Seg} IMF, often show signatures of two separate runaways, one weaker than the 
other.  These runaways may produce two VMSs, both of which may create massive black 
holes.  If the VMSs do not collide prior to creating the massive black holes a massive binary 
black hole system can be formed.  If produced, an IMBH-IMBH binary is a good 
source for LISA during the inspiral phase \citep[see][]{2006ApJ...640L..39G}.  
\subsection{Enhanced Wind Mass Loss: Helium Enrichment}
\label{wind}
If the wind mass loss of the most massive star dominates over its
collisional growth, then the rapid increase in mass is inhibited. (Figure\ 
\ref{plot:wind}).  In this case a VMS is not produced as a result of the 
runaway. Instead, a large amount of processed stellar material is injected into 
the surrounding interstellar medium, which is, consequently, expected to become 
strongly enriched in Helium. For example, the total mass lost via winds from the 
most massive star formed in the runaway in the run {\tt SW8-2} is $\sim 
10^3\,\rm{M_\odot}$, while its $X_{He,S}$ can exceed $0.9$, 
resulting in $\approx 500\,\rm{M_\odot}$ of Helium ejected via winds.
In contrast, if the stars, that merged to form the VMS, are evolved in isolation, 
the total mass of ejected Helium would be significantly lower, $\approx 100\,\rm{M_\odot}$, 
within the first $3\,\rm{Myr}$.  
An anomalously high Helium enrichment of the surrounding interstellar medium 
may, therefore, provide an observational signature for a collisional runaway. 

However, such a signature might not necessarily be found inside Arches but
rather outside of it, as the material lost through winds has rather large
velocities. Assuming the terminal flow velocity of the wind from O/B stars for
solar metallicities is $0.5$ times the escape speed from the surface of the
star \citep{2001A&A...369..574V}, we find that the wind velocity for our collisionally 
formed VMSs is $\approx 500\,\rm{km/s}$.   
As these velocities are much larger than the escape speed of even much more
massive star clusters, it is likely that the material escapes the Arches
cluster entirely.
On the other hand, this material can also slow down through collisions with
winds from other massive stars. Such multiple stellar wind interactions result
in a complex network of shock compressed, extremely hot ($10^7-10^8\, K$) gas
that should then leave the cluster as a cluster wind \citep[see,
e.g.,][]{0004-637X-536-2-896}. The structure of this wind consists of four
zones \citep{2004ApJ...610..226S}: a star cluster region filled with a hot
X-ray plasma, an X-ray halo with decreasing temperature profile, a
line-cooling, recombining zone, and a region of recombined gas, exposed to the
UV radiation from the central star cluster. Indeed diffuse X-ray bright
sources have been observed within Arches as well as extending beyond the
cluster radius \citep[e.g.][]{2002ApJ...570..665Y,2005ApJ...623..171R}. This
extremely hot gas may also be confined due to ram pressure exerted by a
molecular cloud or other external medium surrounding the cluster
\citep{2005ApJ...623..171R}.

Helium abundances are usually determined through measuring Helium recombination 
lines \citep[e.g.,][]{2000ApJ...541..688P}.  Thus a strong 
Helium enrichment
can then only be detected outside the X-ray halo region, in the line cooling
zone. Based on detailed calculations of \citet{2004ApJ...610..226S}, this region is
outside a radius of $3$-$7\,\rm{pc}$ from the Arches cluster.
Other signatures of a collisional runaway might be provided by the wind structure 
itself.  This is because the extension and physical parameters of the
diffuse X-ray emitting region are mainly dependent on the sum of the mass-loss
rates of the individual sources. Given that for the case of a runaway this
mass loss rate can be significantly increased, it may strongly influence the 
appearance of the cluster winds.  
However, such imprints might be very diffucult to detect. This is in part 
because our rates for the most massive star are highly time-variable, varying 
by two orders of magnitudes within a few $10^5\,\rm{yr}$, while particularly 
large values in excess of the total mass injection rate for all other Arches 
stars of $7\approx \times 10^{-4}\,\rm{M}_\odot\,\rm{yr}^{-1}$ 
\citep{2003MNRAS.339..280S} are only attained during an even shorter timescale.  
Another reason is that the Arches cluster wind may be confined by cooler 
surrounding inter stellar material, an effect that is hard to quantify 
\citep{2005ApJ...623..171R}.

Instead of searching for imprints of a runaway in the global cluster wind 
structure, signatures may be found in the properties of individual objects, as 
stellar radii and mass loss rates of runaway stars can be unusual 
\citep[Figure\ \ref{plot:wind};][]{2009A&A...497..255G}.
One way to directly detect and constrain the mass loss rate, the effective 
radius of the star, and the wind velocities is by exploiting the so called Baldwin 
effect \citep{1977ApJ...214..679B,1993ApJ...414L..25M}, which is an 
anticorrelation between the line luminosity of one strong emission line, such 
as HeII (often found in WR winds), and the continuum luminosity.  
\citet{2001A&A...372..963V} show that the ratio of line and continuum 
luminosity, the equivalent width $W$, depends on the mass loss rate, $\dot{M}$, 
the terminal velocity of the wind, $v_\infty$, and the effective radius , 
$R_\star$, as $W \propto \dot{M}^2 v_\infty^{-2} R_\star^{-3}$. When the 
distance of the star is known, and the temperature can be estimated, 
observation of $W$ can constrain the wind mass loss rate, wind velocities and 
the effective radii of the stars.  

\section{Summary and Conclusion}
\label{conclusion}
We carried out many $N$-body simulations with initial conditions chosen such as to
resemble the Arches cluster. Using a standard \citet{2001MNRAS.322..231K} IMF 
with no radial variation ({\tt NSeg}) and a radially dependent IMF that corresponds to 
the level of mass segregation inferred from observations 
\citep[{Seg}; e.g.][]{2007MNRAS.381L..40D}, 
we study possible consequences and discuss potential observable signatures to 
distinguish between the two cases.

We show 
that the degree of mass segregation via dynamical evolution of 
the simulated clusters during the first $2\,\rm{Myr}$ is minor both for the {\tt Seg} and 
the {\tt NSeg} IMF (Figure\ \ref{plot:MF}).  Thus, while runs with the {\tt Seg} 
IMF reproduce well all parts of the currently observed slopes in the MF at $\approx 
2\,\rm{Myr}$, those with the {\tt NSeg} IMF do not produce enough mass 
segregation within the Arches lifetime to match the observed MF as well 
(Figure\ \ref{plot:MF}).  Therefore, our results indicate that the present day observed 
degree of mass segregation in the Arches cluster cannot be a result of dynamics 
only and at least some degree of primordial mass segregation is required.

We further show that the choice of IMF changes the overall dynamical evolution significantly 
for Arches like clusters over a range of initial $W_0$.  For the {\tt Seg} IMF 
there is clear evidence of collisional runaway, producing stars up to $\sim 
10^3\ \rm{M_\odot}$ over a range of initial concentrations (Table\ 
\ref{tab:runs}, Figures\ \ref{plot:tvsm_w8-1}, \ref{plot:tvsm_w7-1}, 
\ref{plot:wvsm}).  In contrast, with the more standard {\tt NSeg} IMF 
no collisional runaway is observed with the same initial conditions.  
We find 
that the sequence of collisions in the simulations with successful collisional 
runaway starts at around $2\,\rm{Myr}$.  Note that the current estimated age of 
the Arches is $2 \pm 1\,\rm{Myr}$ \citep[e.g., ][]{2002A&A...394..459S}.  
Hence, if the observed top-heavy MF within $r \sim 2r_c$ is indeed primordial, 
then it is possible that a collisional runaway has started or is bound to 
happen.  Unfortunately, the uncertainty in the age estimation prevents us from 
predicting whether a collisional runaway should already have taken place in the 
Arches.      

An interesting aspect of the collisional runaways observed in this study is the general 
tendency of a double runaway in cases where the collisional runaway is strong (e.g., runs 
{\tt SW8-1,2} and runs {\tt SW7-1,2}, see Table\ \ref{tab:runs}).  
Double collisional runaways have been observed 
and studied in numerical simulations before \citep{2006ApJ...640L..39G}.  However, those 
simulations concluded that primordial binaries are a necessity for a double collisional 
runaway.  It is very interesting that with the primordially mass-segregated clusters even without 
any primordial binaries double collisional runaways can take place.  Due to the flatness in 
the high mass end of the IMF in this case, there is a larger reservoir of high-mass stars.  
Thus the probability to start a second runaway is higher.  
The ultimate fate of the smaller runaway is dependent on the statistical fluctuations of 
the simulation (see \S\ref{results}).  However, it is possible that the 
VMSs do not collide prior to compact object formation and both stay bound to the 
cluster after their respective supernova explosions (e.g., run {\tt SW8-1}).  
In such a case they may form IMBH binaries or coalesce at the center of the cluster.  
If possible to form, this can create a very strong gravity wave signal for LISA \citep{2006ApJ...640L..39G}.  

If the mass loss from stellar evolution driven winds is too high then the 
collisional mass growth is quenched \citep{2009A&A...497..255G}.  By 
evolving the collision products separately following each successive collision 
in a sequence of runaway collisions according to the prescription in 
\citet{2009A&A...497..255G} we find that the VMSs are not produced as a result 
of a collisional runaway.  Instead, a Helium star of a few tens of a Solar mass 
is created (Figure\ \ref{plot:wind}).  The mass lost via these winds is 
extremely Helium rich, with $X_{He, S}$ reaching beyond $0.9$.  In this case 
the cluster as well as its surrounding medium will be enriched by this Helium 
rich gas, which could be observed through recombination at distances beyond 
$3-7\,\rm{pc}$ from the Arches cluster.

Further indications of a collisional runaway might be provided by the runaway 
object itself, as such an object has a rather unusual radius and mass loss rate 
compared to ordinary massive stars. Constraints on the radius, mass loss rate, 
and the wind speeds may be obtained using the Baldwin effect 
\citep[e.g.,][]{1977ApJ...214..679B,2001A&A...372..963V}.  Furthermore, during the 
evolution of the runaway star, a Pistol like stellar object may also be produced 
(\S\ref{pistol}).  However, since the duration for which the star attains Pistol like high 
luminosities is short, it is not possible to predict whether a Pistol like star should 
already have been created in Arches given the uncertainties in the age estimation 
and the statistical fluctuations in the exact onset and the details of the simulated 
collisional runaways.      

Nevertheless, we should point out that the above recipe to follow the runaway 
star for these high stellar winds is not completely 
self-consistent.  Each collision in a cluster is dependent on the masses and 
radii of the collision progenitors.  Since, severe wind mass loss from the 
stars changes the masses of the stars significantly, the chain of collisions in 
the runaway may also be different.  
However, this model still provides us with an estimate of the total
mass loss from the VMS via stellar winds, its stellar properties, as well as the 
composition of the ejected material if a runaway did indeed happen.  

We thank Farhad Yusef-Zadeh for helpful suggestions.  This research was 
supported by NASA Grant NNX08AG66G and NSF Grant AST-0607498 at Northwestern University.  This research was partly done at KITP 
while the authors participated in the spring 2009 program on ``Formation and 
Evolution of Globular Clusters", and was supported in part by the NSF Grant 
PHY05-51164. 

\bibliography{biblio_arches}

\end{document}